# Phase stability, elastic, electronic, thermal and optical properties of Ti$_3$Al$_{1-x}$Si$_x$C$_2$ (0 ≤ $x$ ≤ 1): First principle study

**M.S. Ali, A.K.M.A. Islam[*], M.M. Hossain, F. Parvin**

*Department of Physics, Rajshahi University, Bangladesh*

ABSTRACT

The structural parameters with stability upon Si incorporation and elastic, electronic, thermodynamic and optical properties of Ti$_3$Al$_{1-x}$Si$_x$C$_2$ (0 ≤ $x$ ≤ 1) are investigated systematically by the plane wave pseudopotential method based on the density functional theory (DFT). The increase of some elastic parameters with increasing Si-content renders the alloys to possess higher compressive and tensile strength. The Vickers hardness value obtained with the help of Mulliken population analysis increases as $x$ is increased from 0 to 1. The solid solutions considered are all metallic with valence and conduction bands, which have a mainly Ti 3d character, crossing the Fermi level. The temperature and pressure dependences of bulk modulus, normalized volume, specific heats, thermal expansion coefficient, and Debye temperature are all obtained through the quasi-harmonic Debye model with phononic effects for $T = 0–1000$ K and $P = 0–50$ GPa. The obtained results are compared with other results where available. Further an analysis of optical functions for two polarization vectors reveals that the reflectivity is high in the visible-ultraviolet region up to ~10.5 eV region showing promise as good coating material.

## 1. Introduction

The ternary layered ceramic compound Ti$_3$AlC$_2$, a member of the well-known MAX phases, exhibits unique properties combining the merits of metals and ceramics. The compound is characterized by its high melting point, low density, excellent oxidation resistance, high strength at high temperatures, significant ductility, good electrical conductivity, and having self-lubrication and good machinability [1-19]. These remarkable properties render Ti$_3$AlC$_2$ a technologically interesting material and hold promise for applications in high temperature applications. As a result Ti$_3$AlC$_2$ and its isostructural counterpart Ti$_3$SiC$_2$ have been subjected to both experimental and theoretical investigations by many researchers [20-43] to highlight their characteristics. Some of these results are highlighted in two recent review works by Wang and Zhou [19] on layered machinable and electrically conductive Ti$_2$AlC and Ti$_3$AlC$_2$, and by Barsoum and Radovic [20] on the current understanding of the elastic and mechanical properties of bulk MAX phases.

The two MAX phases Ti$_3$AlC$_2$ and Ti$_3$SiC$_2$ are isostructural and hence these can form solid solutions that may possess properties of both these phases. Xu et al. [25] studied electronic and bonding properties of the solid solution Ti$_3$Si$_{1-x}$Al$_x$C$_2$ by first-principle calculations and the results demonstrated that as Al-content is increased, all the bonds weakened to certain extents that lead to an unstable structure both energetically and geometrically. Wang and Zhou [32] carried out a first-principles study of Ti$_3$Si$_{0.75}$Al$_{0.25}$C$_2$ solid solution to investigate the changes in the equilibrium properties and electronic structure of Ti$_3$SiC$_2$. Previous theoretical and experimental works show that it is possible to increase the hardness and strength of Ti$_3$AlC$_2$ by alloying. Zhou et al. [18] synthesized a series of Ti$_3$Al$_{1-x}$Si$_x$C$_2$ ($x$ ≤ 0.25) solid solutions using an in situ hot pressing/solid-liquid reaction method. They observed a significant strengthening of $Ti_3AlC_2$ by incorporation of Si to form $Ti_3Al_{1-x}Si_xC_2$ solid solutions. However they limited their study on the strengthening aspects only for $x \leq 0.25$.

[*] Corresponding author. Tel.: +88 0721 750980; fax: +88 0721 750064.
E-mail address: azi46@ru.ac.bd (A.K.M.A. Islam).

We know that many macroscopic properties of MAX phases are closely related to the electronic structure of the compound. Investigation of the electronic structure and bonding properties is thus essential for understanding the various properties of $Ti_3AlC_2$. Zhou et al. [21] calculated the electronic structure of $Ti_3AlC_2$ using the ab initio total-energy pseudopotential method based on the density functional theory, and analyzed its chemical bonding properties. It is necessary to see the effects of the influence of incorporation of different amounts of Si on the physico-chemical and electronic properties for the entire range. It is well-known that the thermodynamic properties are the basis of solid state science and industrial applications since they reveal the specific behavior of materials under high pressure and high temperature environments. Thus knowledge of the heat capacity of a substance not only provides essential insight into its vibrational properties but also is necessary for many applications. Although some thermodynamic properties of $Ti_3AlC_2$ have been studied theoretically [31] and experimentally [30], a thorough theoretical investigation for the entire stoichiometric range will be useful to provide a deep understanding of their thermodynamic properties.

We note that despite comprehensive knowledge on the properties of the stoichiometric compound, less information is available for the solid solution $Ti_3Al_{1-x}Si_xC_2$. In the present work, the focus will be on areas where little or no work has been carried out, particularly for the solid solution in the covering entire doping range. The phase stability, change of structural parameters (e.g., lattice parameters), mechanical and electronic properties with Si content in $Ti_3Al_{1-x}Si_xC_2$ ($0 \leq x \leq 1$) will be investigated. The Vickers hardness will be estimated for $x = 0, 1$ and discussed. We will also carry out a comparative study on the thermodynamic properties of $Ti_3Al_{1-x}Si_xC_2$ ($0 \leq x \leq 1$) using the quasi-harmonic approximation within the density functional theory (DFT). Further the parameters of optical properties (dielectric function, absorption spectrum, conductivity, energy-loss spectrum and reflectivity) will be calculated and discussed.

## 2. Computational methods

The zero-temperature energy calculations have been performed using CASTEP code [44] which utilizes the plane-wave pseudopotential based on density functional theory (DFT). The electronic exchange-correlation energy is treated under the generalized gradient approximation (GGA) in the scheme of Perdew-Burke-Ernzerhof (PBE) [45]. The interactions between ion and electron are represented by ultrasoft Vanderbilt-type pseudopotentials for Ti, Al, Si and C atoms [46]. The crystal models of solid solution $Ti_3Al_{1-x}Si_xC_2$ ($x = 0, 0.25, 0.5, 0.75, 1$) have been built using the supercell method based on the structure of $Ti_3AlC_2$. The 24-atom supercells are built as two times of the unit cell of $Ti_3AlC_2$ for the case of $x = 0.5$ and four times for $x = 0.25, 0.75$ (48-atom supercell). The calculations use a plane-wave cutoff energy 500 eV for all cases. For the sampling of the Brillouin zone, $10 \times 10 \times 2$ $k$-point grids generated according to the Monkhorst-Pack scheme [47] are utilized. These parameters are found to be sufficient to lead to convergence of total energy and geometrical configuration. Geometry optimization is achieved using convergence thresholds of $5 \times 10^{-6}$ eV/atom for the total energy, 0.01 eV/Å for the maximum force, 0.02 GPa for the maximum stress and $5 \times 10^{-4}$ Å for maximum displacement. Integrations in the reciprocal space were performed by using the tetrahedron method with a $k$-mesh of 54 $k$-points in the irreducible wedge of Brillouin zone (BZ). The total energy is converged to within 0.1mRy/unit cell during the self-consistency cycle.

To investigate the thermodynamic properties, we employed the quasi-harmonic Debye model, the detailed description of which can be found in the literatures [48, 49]. Through this model, one could calculate the thermodynamic parameters including the bulk modulus, thermal expansion coefficient, specific heats, and Debye temperature, etc. at any temperatures and pressures using the DFT calculated $E$-$V$ data at $T = 0$ K, $P = 0$ GPa and the third-order Birch-Murnaghan EOS [50].

## 3. Results and discussion

*3.1. Structural parameters and stability upon Si incorporation*

Ti$_3$AlC$_2$ has two polymorphs in α- and β-form, but the former has a lower total energy (− 0.22 eV/f.u.) than the latter-type and is energetically more favorable. We will thus consider the α-polymorph Ti$_3$AlC$_2$ which crystallizes in a hexagonal structure with a space group of P6$_3$/mmc, with Ti(1) atoms in position 2*a*, and Ti(2) atoms in 4*f* (*z*=0.129), Al atoms in 2*b*, C atoms in 4*f* (*z*=0.570). There are two formula units per unit cell. The phase stability of the energetically favorable α-Ti$_3$SiC$_2$ has also been studied [22] and the estimate shows that *α→β* transition will take place at *P* ~380-400 GPa. To investigate the ground-state properties of Ti$_3$Al$_{1-x}$Si$_x$C$_2$ we obtained the structural configurations that reached the energy and force convergence. The calculated fully relaxed equilibrium values of the structural parameters of Ti$_3$Al$_{1-x}$Si$_x$C$_2$ (0 ≤ *x* ≤ 1) are presented in Fig. 1 together with other available data [17, 18, 28, 43]. It is evident from the figure that the lattice parameters change as Si is incorporated in Ti$_3$AlC$_2$. The dramatic decrease of lattice parameter *c* as a function of Si content is borne out satisfactorily by the available measured values for *x* ≤ 0.25 and *x* = 1 [17, 18, 43]. On the other hand theoretical *a*, in excellent agreement with the available measured values [17, 18, 43], remained almost unchanged for the entire ranges of *x* studied. The volume of the solid solution decreases linearly with increasing *x*. From the average nearest-neighbor bond lengths of Ti-C, Ti-Si, and Ti-Al, it is seen that the change in bond length of Ti-C, and Ti-Si (*x*=1) bonds is small whereas the bond length of Ti-Al (*x*=0) decreases upon increasing Si content, in total by ~0.2 Å. However, the hexagonal structure of the material is retained.

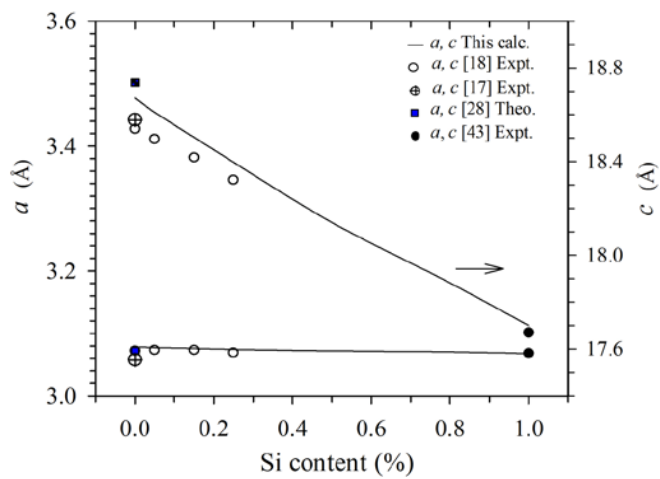

**Fig. 1.** Lattice parameters *a* and *c* as a function of Si-content in Ti$_3$Al$_{1-x}$Si$_x$C$_2$.

Previous study showed that the solid solution can be formed in the Ti$_3$AlC$_2$-Ti$_3$SiC$_2$ system, with experimental observations of strong indications of Si substituting for Al for *x* ≤ 0.25 [18]. It may be useful to discuss here the phase equilibria of the solid solutions Ti$_3$Al$_{1-x}$Si$_x$C$_2$. The strength of the forces that bind atoms together in the solid state is measured by its cohesive energy which is relevant in studying the phase equilibrium. The cohesive energy of Ti$_3$Al$_{1-x}$Si$_x$C$_2$ per atom is defined as the total energy of the constituent atoms at infinite separation minus the total energy of the compound (per f.u.) at equilibrium configuration [32, 51]. The cohesive energy of Ti$_3$AlC$_2$ is found to be close but less than that of Ti$_3$SiC$_2$ i.e. $E_{coh}^{x=0} < E_{coh}^{x=1}$. The calculated cohesive energies of Ti$_3$Al$_{1-x}$Si$_x$C$_2$ for *x* = 0.25, 0.5, and 0.75 are all between those of the two end members. This indicates that the solid solutions Ti$_3$Al$_{1-x}$Si$_x$C$_2$ can be formed from the viewpoint of energy favorability.

*3.2. Elastic properties*

It is now well established that first principle studies based on the density-functional theory can be used to obtain reliable elastic properties of inorganic compounds. The elastic constant tensor of $Ti_3Al_{1-x}Si_xC_2$ ($x$ = 0 to 1.0) is reported in Table 1 along with available computed elastic constants of $Ti_3AlC_2$ and $Ti_3SiC_2$ [22, 26, 41, 53]. Using the second order elastic constsnts, the bulk modulus $B$, shear modulus $G$, and Young's moduluds $Y$ (all in GPa), and Poisson's ratio $v$ at zero pressure are calculated and illustrated in the table. The pressure dependence of the elastic constants is a very important characterization of the crystals with varying pressure and/or temperature, but we defer it till in a later section. It is seen that $Ti_3Al_{1-x}Si_xC_2$ are characterized by high elastic constants, as evident from the obtained values of Young's modulus and bulk modulus.

**Table 1**

Elastic constants $C_{ij}$, the bulk modulus $B$, shear modulus $G$, Young's modulus $Y$ (all in GPa) and Poisson's ratio $v$ and linear compressibility ratio $k_c/k_a$, at $T$ = 0 K, $P$ = 0 GPa.

| Phase | $C_{11}$ | $C_{12}$ | $C_{13}$ | $C_{33}$ | $C_{44}$ | $B$ | $G$ | $Y$ | $v$ | $A$ | $k_c/k_a$ |
|---|---|---|---|---|---|---|---|---|---|---|---|
| $Ti_3AlC_2$[a] | 361 | 75 | 70 | 299 | 124 | 160 | 131 | 309 | 0.18 | 0.87 | 1.29 |
| $Ti_3AlC_2$[b] | 353 | 75 | 69 | 296 | 119 | 159 | 128 | 302 | 0.183 | 0.86 | 1.28 |
| $Ti_3AlC_2$[c] | - | - | - | - | - | 165 | 124 | 297 | 0.20 | - | - |
| $Ti_3AlC_2$[d] | 368 | 81 | 76 | 313 | 130 | 168 | 135 | 320 | 0.18 | 0.90 | 1.25 |
| $Ti_3AlC_2$ | 354 | 73 | 66 | 297 | 122 | 157 | 130 | 305 | 0.163 | 0.87 | 1.28 |
| $Ti_3Al_{0.75}Si_{0.25}C_2$ | 361 | 75 | 77 | 318 | 138 | 166 | 138 | 324 | 0.175 | 0.96 | 1.17 |
| $Ti_3Al_{0.50}Si_{0.50}C_2$ | 365 | 82 | 84 | 338 | 146 | 174 | 141 | 333 | 0.181 | 1.03 | 1.10 |
| $Ti_3Al_{0.25}Si_{0.75}C_2$ | 368 | 84 | 90 | 353 | 153 | 180 | 144 | 341 | 0.183 | 1.08 | 1.03 |
| $Ti_3SiC_2$ | 367 | 86 | 96 | 351 | 153 | 182 | 143 | 340 | 0.189 | 1.084 | 1.02 |
| $Ti_3SiC_2$[e] | 366 | 94 | 100 | 352 | 153 | 186 | 141 | 338 | 0.20 | 1.125 | 1.03 |
| $Ti_3SiC_2$[c*] | - | - | - | - | - | 185 | 139 | 333 | 0.20 | - | - |
| $Ti_3SiC_2$[f] | - | - | - | - | - | 185.6 | 143.8 | 343 | 0.192 | - | - |

[a] [22]

[b] [26]

[c] [34] Experimental values for $Ti_3Al_{1.1}C_{1.8}$ sample at room temperature.

[d] [41]

[e] [53]

[f] [34] Experimental values for fine-grained samples at room temperature.

[g] [52] Experimental (RUS) at 298 K.

The ductility of a material can be roughly estimated by the ability of performing shear deformation, such as the value of shear-modulus-to-bulk-modulus ratios. Thus a ductile plastic solid would show low $G/B$ ratio ($< 0.5$); otherwise, the material is brittle. As is evident from Table 1, the calculated $G/B$ ratios (0.82 – 0.78) for $Ti_3Al_{1-x}Si_xC_2$, which decrease slightly with $x$, are comparable to those of other 312 MAX phases [41]. The elastic anisotropy of crystals, defined by $A = 2C_{44}/(C_{11}–C_{12})$, can be responsible for the development of microcracks in the material (see Ref. [54]). So, the elastic anisotropy of the shear of hexagonal crystals is described by the factor $A$, which is unity for an ideally isotropic crystal. According to our calculations the value of $A$ increases from 0.87 to 1.084 as $x$ increases from 0 to 1. Yet another anisotropy parameter that can be estimated as the ratio between the uniaxial compression values along the $c$ to that of $a$ axis for a hexagonal crystal: $k_c/k_a = (C_{11}+C_{12}–2C_{13})/(C_{33}–C_{13})$. The calculated data of Table 1 for $Ti_3Al_{1-x}Si_xC_2$ show that the crystal compression along the $c$ axis is larger than that along the $a$ axis and it decreases with the value of Si-content $x$.

*3.3. Electronic structure and bonding characteristics*

The calculated energy band structure and density of states (DOS) of $Ti_3Al_{1-x}Si_xC_2$ are shown in Figs. 2 and 3, respectively. Although we have not included the results for $Ti_3SiC_2$ ($x = 1$) in the figure, these will nevertheless be discussed. It may be remarked here that the alloys ($0 \leq x \leq 1$) are all metallic with valence and conduction bands, which have a mainly Ti 3d character, crossing the Fermi level. The values of total DOS at the Fermi level in states/eV are 3.9 (3.76 [32]), 3.54, 2.79, 3.39 (4.09 [32]) and 4.62 (5.04 [32]) for different Si contents under consideration. This difference between our values and those available from Wang and Zhou [32] may be attributed to the different scheme LDA used by these authors for treating the electronic exchange correlation energy.

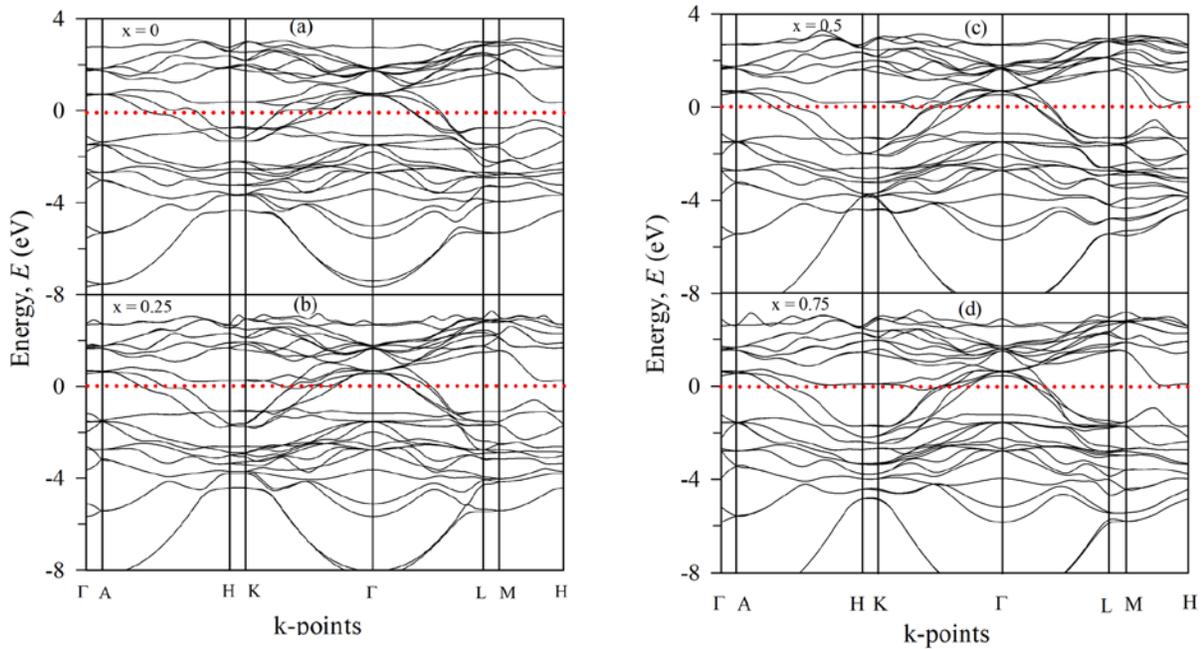

**Fig. 2.** Band structures of $Ti_3Al_{1-x}Si_xC_2$ for (a) $x = 0$, (b) $x = 0.25$, (c) $x = 0.5$ and, (d) $x = 0.75$.

Fig. 3 displays the partial density of states (PDOS) of each element for different Si contents ($0 \leq x \leq 0.75$). The atomic bonding characteristics are clearly illustrated in these PDOS. C does not contribute significantly to the DOS at the Fermi energy and therefore is not involved in electronic transport. The same can be said about the Al and Si atoms. The lowest valence bands of the PDOS are comprised of C 2p states with a small mixture of Ti 3p/3d, Al 3s/3p, Si 3s/3p states between −8.4 (data shown from −8 eV) to 0eV. The bonding peak located around −2.6 eV corresponds to the strong hybridization of Ti 3d and C 2p states. As can be seen in Table 2 the calculated $Ti_1$-C bond length (2.075 Å) in $Ti_3AlC_2$ is ~ 5.6% shorter than that of Ti2-C (2.198 Å). The corresponding $Ti_1$-C bond length (2.0891 Å) in $Ti_3SiC_2$ ($x=1$) is 4.5% shorter than 2.198 Å for $Ti_2$-C bond. Similar results have been reported by Barsoum et al. [55] for $Ti_3SiC_2$. This implies that the $Ti_1$-C bond is stronger than that of $Ti_2$-C (see also bond population analysis, Table 2). The bonding peak, at about −1 eV, relates to the $Ti_1$ 3d and Al 3p/Si 3p hybridization states. These states appear at a higher energy range than those of the Ti-C d-p states. The Ti-Al/Si bond can be considered to be relatively weaker than the Ti-C covalent bond.

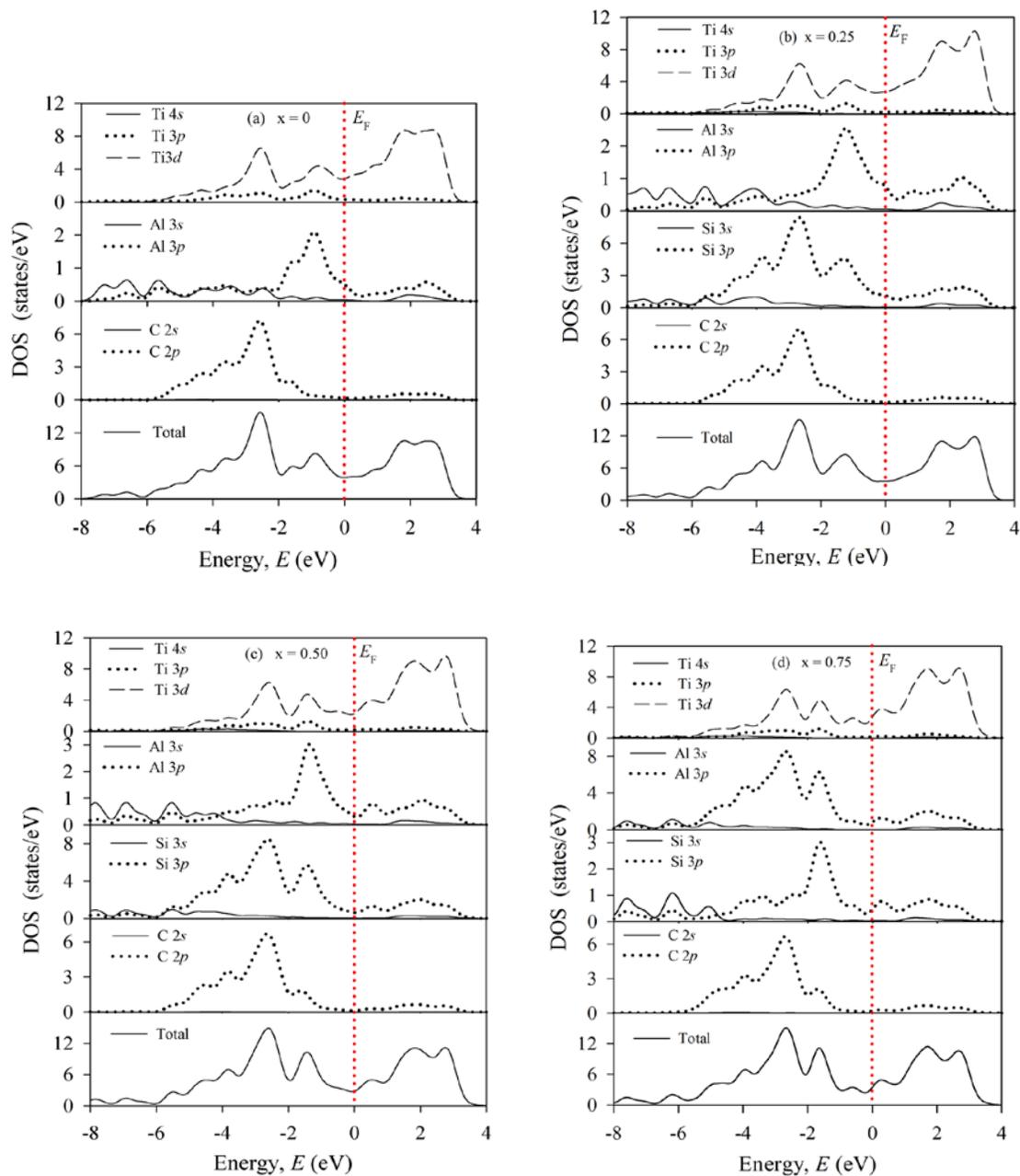

**Fig. 3.** Total and partial density of states of (a) $x = 0$, (b) $x = 0.25$, (c) $x = 0.50$, and (d) $x = 0.75$ for $Ti_3Al_{1-x}Si_xC_2$.

*3.4. Mulliken bonding population and theoretical hardness*

The Mulliken bond populations are calculated to understand the bonding behavior as well as to obtain Vickers hardness ($H_V$) of $Ti_3AlC_2$ and $Ti_3SiC_2$. The relevant formula for the hardness is given as [35, 36]:

$$H_V = \left[ \prod^\mu \left\{ 740\left(P^\mu - P^{\mu'}\right)\left(v_b^\mu\right)^{-5/3} \right\}^{n^\mu} \right]^{1/\Sigma n^\mu} \tag{1}$$

where $P^\mu$ is the Mulliken population of the μ-type bond, $P^{\mu'} = n_{\text{free}}/V$ is the metallic population and $v_b^\mu$ is the bond volume of μ-type bond.

The calculated results are given in Table 2. The Mulliken bond populations explain the overlap degree of the electron clouds of two bonding atoms. Hence the highest value indicates the strong covalency of the chemical bonding. Comparing the bond overlap population $P^\mu$ of $Ti_3AlC_2$ and $Ti_3SiC_2$ we find that the value of $P^\mu$ for $Ti_1$-Si bond is about 1.3 times larger than that of $Ti_1$-Al bond. In other words, it can be said that $Ti_1$-Si bond is 30% stronger than $Ti_1$-Al bond.

**Table 2**

Calculated Mulliken bond overlap population of μ-type bond $P^\mu$, bond length $d^\mu$, metallic population $P^{\mu'}$, bond volume $v_b^\mu$ (Å$^3$) and Vickers hardness of μ-type bond $H_v^\mu$ and $H_v$ of $Ti_3AlC_2$ and $Ti_3SiC_2$.

| Compound | Bond | $d^\mu$ (Å) | $P^\mu$ | $P^{\mu'}$ | $v_b^\mu$ (Å$^3$) | $H_v^\mu$ (GPa) | $H_v$ (GPa) |
|---|---|---|---|---|---|---|---|
| $Ti_3AlC_2$ | Ti1-C | 2.075 | 1.16 | 0.055 | 7.75 | 26.98 | 9.2 |
| | Ti2-C | 2.198 | 0.87 | 0.055 | 9.21 | 15.08 | |
| | Ti1-Al | 2.9097 | 0.48 | 0.055 | 21.36 | 1.91 | |
| $Ti_3SiC_2$ | Ti1-C | 2.089 | 1.11 | 0.012 | 8.25 | 24.13 | 11.1 |
| | Ti2-C | 2.187 | 0.89 | 0.012 | 9.47 | 15.33 | |
| | Ti1-Si | 2.709 | 0.62 | 0.012 | 17.99 | 3.64 | |

With the help of Mulliken population analysis the Vickers hardness values of $Ti_3Al_{1-x}Si_xC_2$ are shown in Table 2 for only $x = 0$ and 1 as the current methodology does not allow this analysis for disordered systems. Our calculated hardness values are 9.2 GPa for $x = 0$ and 11.1 GPa for $x = 1$. It will be shown in the next section that the bulk modulus $B$ increases with Si-concentration. As $B$ is proportional to the hardness, it is expected that $Ti_3Al_{1-x}Si_xC_2$ will possess higher hardness as $x$ is increased. Indeed we see that the theoretical hardness is higher for $x = 1$ than for $x = 0$. The intrinsic hardness value for $Ti_3AlC_2$, measured by nanoindentation experiments, is 11.4±0.7 GPa [17]. Again, for $Ti_3SiC_2$ thin film, a hardness value of 15 GPa measured by nano-indentation normal to the basal plane has been reported by Molina et al. [37]. Wilhelmsson et al. [38] measured the nanohardness (~15 GPa) for $Ti_3AlC_2$ thin film. Further for $TiO_2$ using the same method, theoretical Vickers hardness has been reported as 11.9 [35] to be compared with the measured value of 12 GPa [39]. The hardness values measured by micro-indentations are in the range of 3−7 GPa for $Ti_3AlC_2$ [6, 18, 20, 40], which are very small compared to the intrinsic hardness value determined by nanoindentation. But it has been remarked by Bei et al. [17] that when indentation tests (micro- or nano-indentation) are performed for large load, several grains are involved in the deformation process. As a result grain boundaries and impurities affect the measured hardness value which is underestimated.

*3.5. Thermodynamic properties*

The elastic parameters and associated physical quantities like Debye temperature etc. allow a deeper understanding of the relationship between the mechanical properties and the electronic and phonon structure of materials. We investigated the thermodynamic properties of $Ti_3Al_{1-x}Si_xC_2$ by using the quasi-harmonic Debye model, the detailed description of which can be found in literatures [48, 49]. Here we computed the normalized volume, bulk modulus, specific heats, Debye temperature and volume thermal expansion coefficient at different temperatures and pressures. For this we utilized *E-V* data obtained from the third-order Birch-Murnaghan equation of state [50] using zero temperature and zero pressure equilibrium values, $E_0$, $V_0$, $B_0$, based on the DFT method as discussed earlier.

Fig. 4 shows the temperature variation of isothermal bulk modulus of $Ti_3Al_{1-x}Si_xC_2$ along with data from Togo et al. [31], the *inset* of which shows bulk modulus $B$ and normalized volume $V/V_0$ data as a function of pressure. Our calculations for different Si content show that $B$ is nearly flat for $T < 150$ K, it then decreases in a slightly non-linear way up to 1000 K. This decrease follows the trend of the volume variation as $T$ increases. For the end member ($x = 0$) our $B$ values and those obtained by Togo et al. [31] are in excellent agreement for the whole temperature range shown. Here, $B$ drops by ~10% from 0K to 1000 K. For $x = 1$, the $B$ values due to Togo et al. [31] is slightly higher at $T = 0$ K, but otherwise the agreement after room temperature is quite good. The inset shows the pressure variation of room temperature $B$ and normalized volume $V/V_0$. Furthermore, it is found that the bulk modulus increases with pressure at a given temperature and decreases with temperature at a given pressure, which is consistent with the trend of volume.

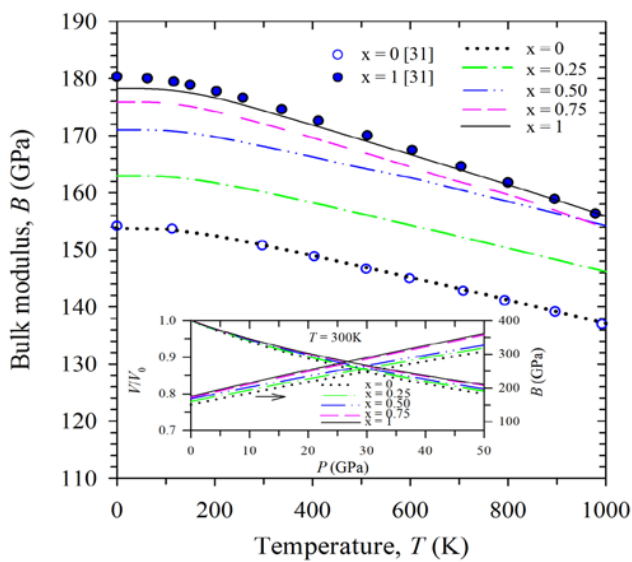
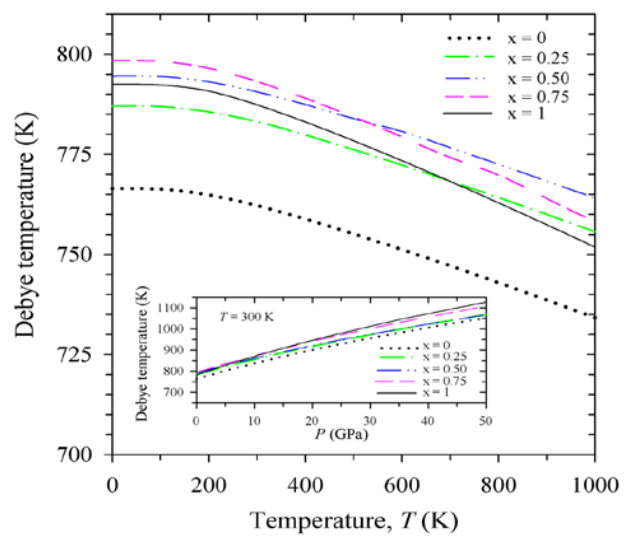

**Fig. 4.** Temperature dependence of $B$ for different Si-contents x. Inset: Pressure dependence of normalized volume and bulk modulus for different Si-contents.

**Fig. 5.** Temperature variation of Debye temperature $\Theta_D$ for different Si-contents x. Inset: Pressure dependence of $\Theta_D$ at different Si-contents.

The temperature dependence of Debye temperature $\Theta_D$ at zero pressure of $Ti_3Al_{1-x}Si_xC_2$ for different Si concentrations is displayed in Fig. 5. It is seen that the $\Theta_D$ for $x = 0$ is the lowest of all other values, but it gradually increases as $x$ is increased. The exception is the Debye temperature for $x = 1$ which is 5 K below the value for $x = 0.75$ at 0 K. We note that the Debye temperature for the solid solutions from $x = 0$ to 1 all decrease non-linearly with temperature. The pressure dependent Debye temperature (presented in the inset at $T = 300$ K) shows a non-linear increase. The variation of $\Theta_D$ with pressure and temperature reveals that the thermal vibration frequency of atoms in $Ti_3Al_{1-x}Si_xC_2$ changes with pressure and temperature. Further for Si content $x = 0$ our calculated Debye temperature of 766 K (at 0 K) is in very good agreement with 760 K determined from elastic measurements [35]. On the other hand for fine grained samples $Ti_3SiC_2$ ($x = 1$), the experimental values 780 K [34] and 715 K [56] may be compared with our estimated value of 781 K at $T = 0$ K.

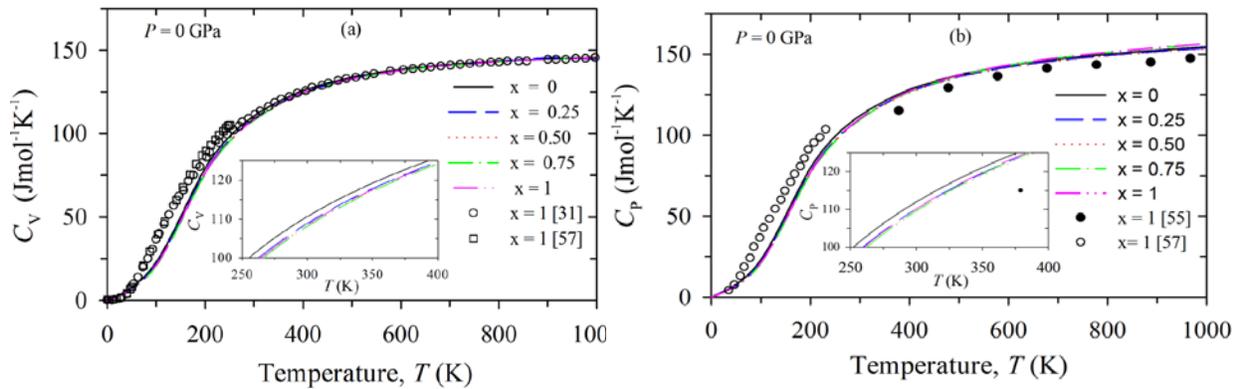

**Fig. 6.** Temperature variation at $P = 0$ GPa of (a) $C_V$, and (b) $C_P$. Other theoretical data on $C_V$ are depicted by squares [57] and open circless [31] for comparison. Measured values of $C_P$ are shown on the right graph by open circles [57] and filled circles [55]. Insets to both figures show magnified curves between 250 and 400 K.

The temperature dependence of constant-volume and constant-pressure specific heat capacities $C_V$, $C_P$ of $Ti_3Al_{1-x}Si_xC_2$ are shown in Fig. 6 (a, b). In fact, the heat capacities increase with increasing temperature, because phonon thermal softening occurs when the temperature increases. The difference between $C_P$ and $C_V$ for $Ti_3Al_{1-x}Si_xC_2$ is given by $C_P - C_V = \alpha_V^2(T)\ BTV$, which is due to the thermal expansion caused by anharmonicity effects. In the low temperature limit, the specific heat exhibits the Debye $T^3$ power-law behavior and at high temperature ($T > 400$ K) the anharmonic effect on heat capacity is suppressed, and $C_V$ approaches the classical asymptotic limit of $C_V = 3nNk_B = 149.6$ J/mol K. These results show the fact that the interactions between ions in $Ti_3Al_{1-x}Si_xC_2$ have large effect on heat capacities especially at low temperatures. As shown in the figure we see that there are very little differences in the $C_V$ values for different Si-contents. Inset shows the magnified graphs for $T$ between 250 K and 400 K. Data on $C_V$ from two different sets of theoretical calculations [31, 57] are also plotted for comparison, which shows that the agreement is quite good. The closeness of the $C_P$ values obtained with different Si-contents is also evident from Fig. 6 (b). The experimental data from two sources [55, 57] are shown for comparison, which indicates that the overall agreement is satisfactory, despite some deviations.

Fig. 7 shows the volume thermal expansion coefficient (VTEC), $\alpha_V$ as a function of (a) temperature and (b) pressure. The thermal expansion coefficient increases rapidly especially at temperature below 300 K, whereas it gradually tends to a slow increase at higher temperatures. On the other hand at a constant temperature, the expansion coefficient decreases strongly with pressure, also in different rate as Si-content is varied. It is well-known that the thermal expansion coefficient is inversely related to the bulk modulus of a material. The measured value of linear thermal expansion coefficients of $Ti_3AlC_2$ and $Ti_3SiC_2$ are $9.2\times10^{-6} K^{-1}$ [31] and $9.3\times10^{-6} K^{-1}$ [31], respectively. Assuming $\alpha_V = 3\times$(linear thermal expansion coefficient), the calculations for these two compounds around room temperature agree well with experiments. As $T$ increases further from room temperature, $\alpha_V$ values also start to increase (rate is higher for $Ti_3SiC_2$). Fig. 7 (b) displays $\alpha_V$ as a function of pressure, which shows a strong pressure-dependent behavior.

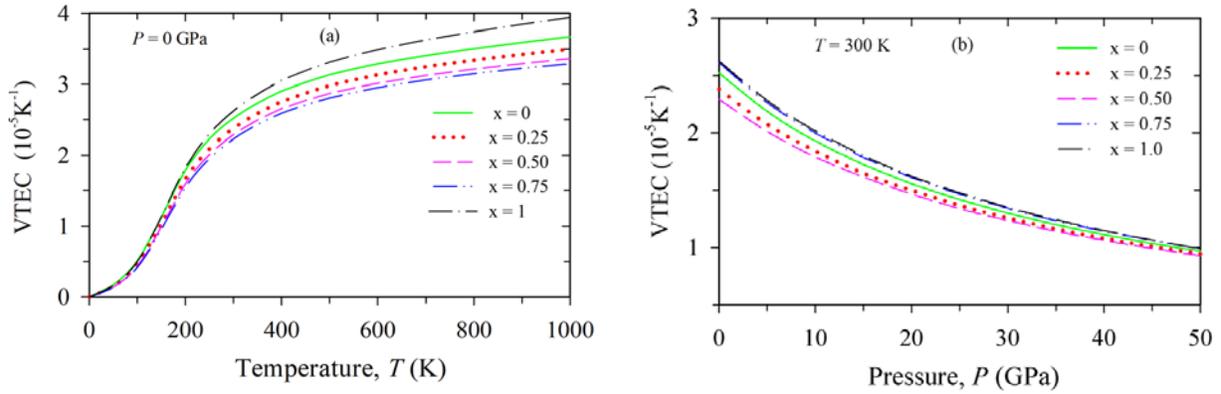

**Fig. 7.** Volume thermal expansion coefficient of $Ti_3Al_{1-x}Si_xC_2$ as a function of (a) temperature for $P = 0$ GPa, and (b) pressure at $T = 300$ K. The variation on Si-content is also displayed.

*3.6. Optical properties*

The analysis of the optical functions of solids helps to give a better understanding of the electronic structure. The complex dielectric function is defined as, $\varepsilon(\omega) = \varepsilon_1(\omega) + i\varepsilon_2(\omega)$. The imaginary part $\varepsilon_2(\omega)$ is obtained from the momentum matrix elements between the occupied and the unoccupied electronic states and calculated directly using [58]:

$$\varepsilon_2(\omega) = \frac{2e^2\pi}{\Omega\varepsilon_0} \sum_{k,v,c} \left|\psi_k^c \left|\mathbf{u}\cdot\mathbf{r}\right|\psi_k^v\right|^2 \delta\left(E_k^c - E_k^v - E\right) \tag{2}$$

where $\mathbf{u}$ is the vector defining the polarization of the incident electric field, $\omega$ is the light frequency, $e$ is the electronic charge and $\psi_k^c$ and $\psi_k^v$ are the conduction and valence band wave functions at $k$, respectively. The real part is derived from the imaginary part $\varepsilon_2(\omega)$ by the Kramers-Kronig transform. All other optical constants, such as refractive index, absorption spectrum, loss-function, reflectivity and conductivity (real part) are those given by Eqs. (49)-(54) in Ref. [58].

Fig. 8 shows the optical functions of $Ti_3Al_{1-x}Si_xC_2$ ($x = 0, 1$) calculated for photon energies up to 20 eV for polarization vectors [100] and [001], along with theoretical spectra of $Ti_3AlC_2$ from Ref. [28]. We have used a 0.5 eV Gaussain smearing for all calculations. This smears out the Fermi level, so that $k$-points will be more effective on the Fermi surface. Despite some variation in heights and positions of peaks, the overall features of the optical spectra of $Ti_3AlC_2$ and $Ti_3SiC_2$ are roughly similar.

Fig. 8 (a) shows the absorption band, similar for both the polarization vectors, in the energy range ~5.1 eV due to its metallic nature. Since the material has no band gap as evident from band structure, the photoconductivity starts with zero photon energy for both polarization vectors as shown in Fig. 8 (b). Moreover, the photoconductivity, and hence electrical conductivity of a material, increases as a result of absorbing photons. There is some variation of the height and width of the peaks between the spectra for $x = 0$ and $x = 1$. Fig. 8 (c) shows the reflectivity spectra (roughly similar for both the polarizations) as a function of photon energy. For polarization vectors [100] and [001], it is found that the reflectivity in Ti$_3$Al$_{1-x}$Si$_x$C$_2$ ($x$ = 0, 1) is high in visible and ultraviolet up to ~10.5 eV region (reaching maximum at 8−9 eV) showing the alloys as promising candidate for use as a coating material.

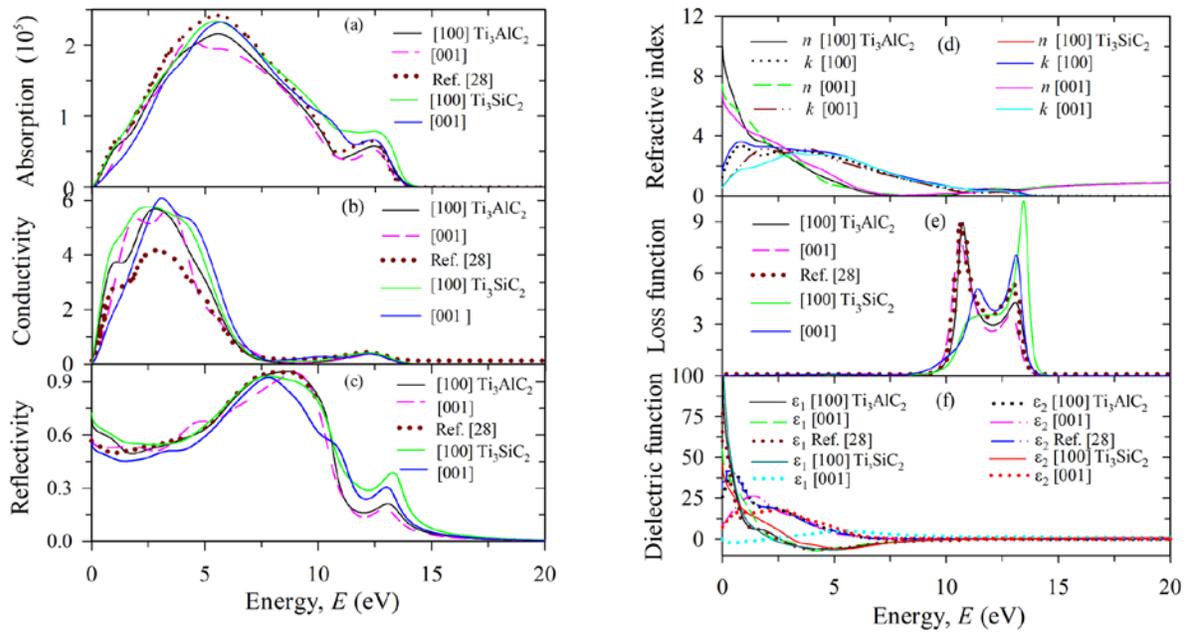

**Fig. 8.** (a) Absorption spectrum, (b) photo-conductivity, (c) reflectivity, (d) refractive index, (e) loss function, and (f) dielectric function of Ti$_3$Al$_{1-x}$Si$_x$C$_2$ for $x$ = 0 and $x$ = 1.

The refractive index and extinction coefficient are illustrated in Fig. 8 (d). The static refractive index of Ti$_3$Al$_{1-x}$Si$_x$C$_2$ is found to have the values 9.5 ($x$ = 0), and 7 ($x$ = 1). In Fig. 8 (e), the electron energy loss function describing the energy loss of a fast electron traversing a material is usually large at the plasma frequency [59]. Prominent peaks are found at 11 eV and 13.1 eV, which indicate rapid reduction in the reflectance. The imaginary and real parts of the dielectric function are displayed in Fig. 8 (f). It is observed that the real part $\varepsilon_1$ vanishes at about 9.5 eV for both $x$ = 0 and 1. Metallic reflectance characteristics are exhibited in the range of $\varepsilon_1 < 0$. The peak of the imaginary part of the dielectric function is related to the electron excitation. For the imaginary part of $\varepsilon_2$, the peak for < 1.5 eV is due to the intraband transitions.

## 4. Conclusion

First-principle calculations based on DFT have been used to study the phase stability, structural, elastic, thermodynamic, electronic and optical properties of the Ti$_3$AlC$_2$ and the effects of Si substitution in it.

The progressively higher calculated values of some elastic parameters with increasing Si-content render the alloys to possess higher compressive and tensile strength. The Vickers hardness value obtained with the help of the Mulliken population analysis increases as $x$ is increased from 0 to 1. The solid solutions considered are all metallic with valence and conduction bands, which have a mainly Ti 3d character, crossing the Fermi level. The atomic bonding characteristics are discussed.

The finite-temperature ($\leq$ 1000 K) and finite-pressure ($\leq$ 50 GPa) thermodynamic properties, e.g. bulk modulus, normalized volume, specific heats, thermal expansion coefficient, and Debye temperature are all obtained through the quasi-harmonic Debye model, which considers the vibrational contribution, and the results are analyzed. The variation of $\varTheta_D$ with temperature and pressure reveals the changeable vibration frequency of the particles in $Ti_3Al_{1-x}Si_xC_2$. The heat capacities increase with increasing temperature, which shows that phonon thermal softening occurs when the temperature increases.

From an analysis of optical functions for both the polarization vectors [100] and [001], it is found that the reflectivity in $Ti_3Al_{1-x}Si_xC_2$ ($x = 0, 1$) is high in visible-ultraviolet region up to ~10.5 eV (reaching maximum between 8 and 9 eV) showing the alloys as promising candidate for use as coating material.